\PassOptionsToPackage{table}{xcolor}
\documentclass[sigconf,9pt,nonacm,balance=false]{acmart}
\usepackage{array}

\newcommand{\placeholder}[1]{\noindent\textit{[Placeholder: #1]}\par}

\newif\ifusebibliography
\usebibliographytrue

\begin{document}

\title[OrderFusion+: Probabilistic Buy--Sell Price Trajectory Forecasting]{OrderFusion+: Probabilistic Buy--Sell Price Trajectory Forecasting in Intraday Electricity Markets}

\author{Runyao Yu\textsuperscript{1,2,3}\qquad Derek W. Bunn\textsuperscript{1}}

\renewcommand{\shortauthors}{Yu and Bunn}

\begin{abstract}
Intraday electricity markets enable participants to adjust energy positions close to delivery, with price forecasts necessary to support trading and the scheduling of flexible electricity resources as well as increasingly responsive consumers. Forecasting model specifications have progressed from using macro-features, such as renewable generation and load, to the micro-features of continuous orderbooks. A recent advanced deep learning model, OrderFusion, explicitly models micro-level buy–sell orderbook interactions. However, despite its superior comparative forecasting performance, it considers only one delivery product at a time, ignoring neighboring-product information. Moreover, when forecasting aggregated price indices such as the liquid German $\mathrm{ID}_3$, $\mathrm{ID}_2$, and $\mathrm{ID}_1$ products, it omits the information contained in price trajectories. In contrast, pretrained time-series foundation models have shown success in financial, renewable-energy, and day-ahead electricity price forecasting. However, their performance on intraday orderbook data remains an open research question of considerable practical importance. In this paper, we propose OrderFusion+, an open-source deep learning model that combines historical orders from the target and neighboring delivery products to forecast probabilistic buy–sell price trajectories. We benchmark OrderFusion+ against forecasting baselines and pretrained foundation models, and investigate dynamic market conditions through the designed dynamic masking mechanism, revealing insights into market efficiency. The implementation and forecasts can be found at: \url{https://runyao-yu.com/OrderFusion/}.

\end{abstract}

\keywords{intraday electricity market, price trajectory forecasting, orderbook, deep learning, neural network}

\maketitle

\begingroup
\renewcommand{\thefootnote}{}
\footnotetext{\textsuperscript{1}London Business School, London, United Kingdom.
\textsuperscript{2}Delft University of Technology, Delft, The Netherlands.
\textsuperscript{3}Austrian Institute of Technology, Vienna, Austria.}
\endgroup

\section{Introduction}
\label{sec:introduction}
Electricity markets coordinate supply and demand through price signals
that guide operational and trading
decisions~\cite{bunn2000forecasting,weron2014review, morstyn2018using}.
In Europe, the main short-term market stages are day-ahead, intraday, and
balancing~\cite{yu2026review}, with delivery products often defined over short intervals, such as 15 minutes.
The day-ahead market establishes energy schedules for the following day.
However, renewable generation and demand remain uncertain as delivery
approaches~\cite{chen2026counterfactual,thakur2023hedging,wang2026adrift}.
Deviations between contracted and actual energy positions are settled at
imbalance prices, potentially exposing participants to substantial
costs~\cite{yu2026mrinn,ganesh2024densities, bruninx2025day}.
Intraday trading allows participants to adjust their day-ahead positions
before delivery as renewable generation forecasts change, reducing their
exposure to imbalance settlement~\cite{yu2026review}.
These opportunities also support repeated position adjustments and
coordinated trading by flexible assets such as
batteries in the continuous intraday
market~\cite{miskiw2025continuous,semmelmann2024rolling, lokhande2022cimtrade}.

Accurate intraday price forecasting can improve continuous intraday trading
decisions~\cite{serafin2022trading,hornek2025value} and support the scheduling
of flexible electricity generation and consumption.
Marcjasz et al.\ use distributional neural networks to forecast intraday
prices and construct prediction bands that determine when to buy or
sell~\cite{marcjasz2023trading}.
Semmelmann et al.\ use XGBoost to predict quantiles of transaction prices
from weather forecast updates and set dynamic buy and sell
thresholds~\cite{semmelmann2026quantile}.
Intraday price forecasts have also been incorporated into rolling optimization
to reschedule electric vehicle charging while meeting users' energy
requirements~\cite{chemudupaty2025ev}.
In water management, intraday price scenarios have been integrated into
stochastic model predictive control to schedule pumping and minimize expected
electricity costs subject to water-level constraints~\cite{heijden2025water}.
Despite this importance, intraday price forecasting remains less extensively
studied than day-ahead forecasting~\cite{yu2026review,narajewski2020econometric}.

For many years, intraday price forecasting methods have incorporated
macro-features such as renewable generation and load.
Examples include Bayesian hierarchical models~\cite{nickelsen2025bayesian},
normalizing flows~\cite{cramer2023flows},
Long Short-Term Memory (LSTM)~\cite{kilic2024lstm}, hierarchical neural network~\cite{thokala2025hierarchical}, and generative neural
networks~\cite{chen2025generative}.
However, these methods often focus on macro-features and aggregate price histories with lower frequency,
without exploring higher-frequency continuous orderbook micro-features that describe
individual buy and sell orders and their interactions.
Furthermore, a study of the German market shows that adding renewable generation
and load forecasts was found not to improve forecasts based on
orderbook-derived price histories, suggesting that the predictive
information in these macro-features was already reflected in the observed prices~\cite{janke2019distribution}.
More recently, 384 orderbook micro-features have been extracted and their
generalization across countries and product types has been
examined~\cite{yu2026orderbookfeatures}.
This line of research led to the development of OrderFusion, an advanced deep learning model designed to explicitly fuse buy and sell orders at the micro-level and produce accurate probabilistic intraday price forecasts~\cite{yu2026orderfusion}.

However, OrderFusion only uses historical trades for one delivery product at a
time, whereas continuous intraday markets allow
multiple delivery products to be traded in parallel.
Modeling dependencies between neighboring
products can improve probabilistic forecasting performance~\cite{hirsch2024crossproduct}.
Moreover, OrderFusion forecasts only a single aggregated price index at a time: $\mathrm{ID}_3$, $\mathrm{ID}_2$, or $\mathrm{ID}_1$. These indices are defined as Volume-Weighted Average Prices (VWAPs) of trades executed within the final three, two, or one hour before delivery, respectively. The final three hours before delivery constitute the most liquid trading window and provide a stable price reference~\cite{epex2020indices}. However, price volatility increases during the final two hours and reaches its highest level in the final hour, as traders face growing imbalance pressure before delivery~\cite{feron2020major}. By compressing each trading window into a single price index, these indices discard the price trajectory and buy--sell dynamics within the window.
Therefore, we make an important development to OrderFusion: we incorporate information from historical trades and neighboring
products through a designed dynamic masking mechanism and shift the forecasting target from price indices to
probabilistic buy--sell price trajectories.

Additionally, foundation models are increasingly being explored for
time-series forecasting, including TimesFM~\cite{das2024timesfm},
Chronos~\cite{ansari2024chronos}, Moirai~\cite{liu2025moirai2}, and TabPFN-TS~\cite{hoo2025tabpfnts}.
These models are motivated by the premise that pretraining deep learning
models on diverse real-world or synthetic datasets can enable strong
zero-shot forecasting on unseen data.
Successful applications have been reported using TimesFM for
financial risk forecasting~\cite{goel2024var}, Chronos for
solar irradiance forecasting~\cite{shan2026solarfm}, and TabPFN for day-ahead
electricity price forecasting~\cite{lipiecki2026foundation}.
However, the performance of these generic foundation models for probabilistic intraday price trajectory forecasting remains unclear. Therefore, we systematically evaluate their probabilistic forecasting performance on commercial intraday orderbook data, which are generally expensive to acquire and have consequently received relatively limited attention in the forecasting literature.
Our contributions are:
\begin{itemize}
  \item We propose and open source OrderFusion+, a novel deep learning model
  that takes historical orders from the target and neighboring delivery
  products as input and produces probabilistic buy--sell price
  trajectory forecasts.
  \item We benchmark OrderFusion+ against a range of baselines. To our
  knowledge, this is the first evaluation of recent 
  foundation models for probabilistic buy--sell price trajectory forecasting
  using commercial intraday orderbook data.
  \item We systematically investigate dynamic market conditions through the designed dynamic masking mechanism, revealing insights into market efficiency.
\end{itemize}

The remainder of this paper is organized as follows.
Section~\ref{sec:preliminary} introduces the continuous intraday market, price indices and trajectories, and OrderFusion.
Section~\ref{sec:model} presents  OrderFusion+.
Section~\ref{sec:baselines} introduces the baselines, and Section~\ref{sec:metrics} defines the evaluation metrics.
Section~\ref{sec:experiment} describes the experimental settings and reports the results.
Section~\ref{sec:conclusion} concludes the paper.

\begin{figure*}[t]
    \centering
    \includegraphics[width=0.96\textwidth]
    {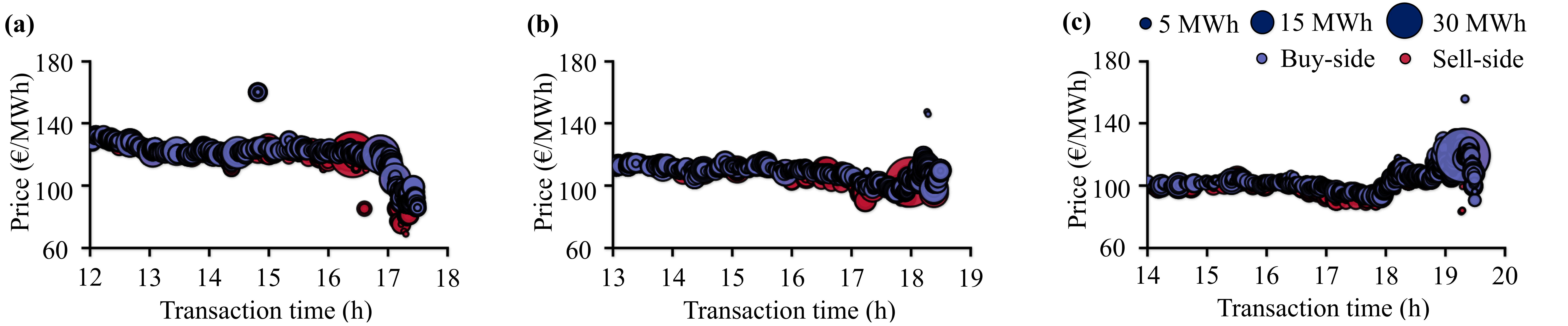}
    \caption{Three example delivery products traded on 23 July 2024 in the German market.
    \textbf{(a)} Delivery at 18:00: the price declines as delivery approaches.
    \textbf{(b)} Delivery at 19:00: the price rises, falls, and rises again as delivery approaches.
    \textbf{(c)} Delivery at 20:00: the price rises and then falls as delivery approaches.
    Traders can submit orders for several delivery products in parallel.}
    \label{fig:neighboring_products}
\end{figure*}

\begin{figure*}[t]
    \centering
    \includegraphics[width=0.96\textwidth]
    {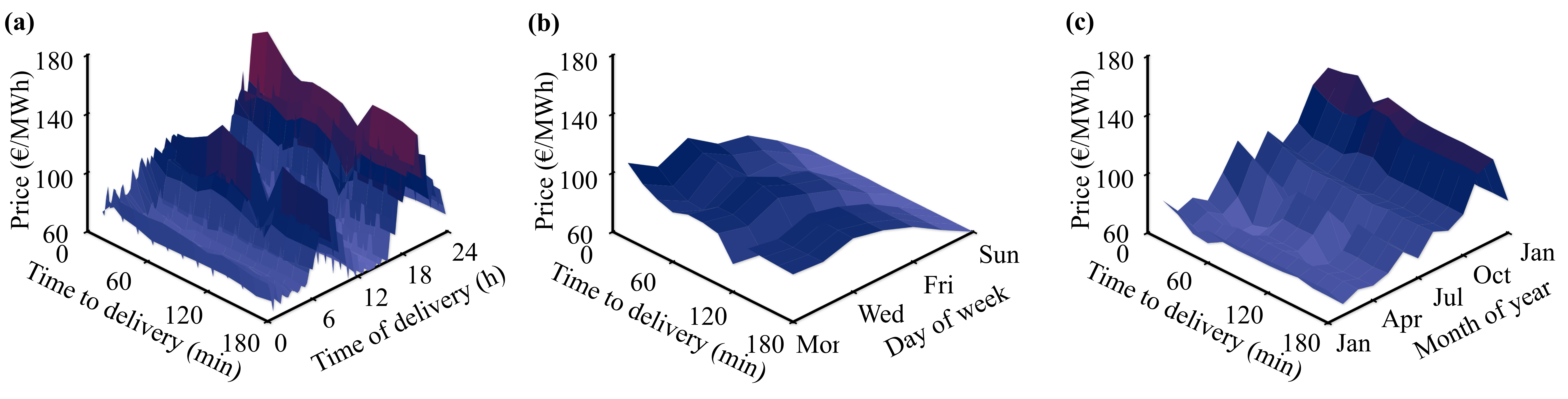}
    \caption{Seasonality of the intraday price trajectory, aggregated over sides during the final 180 minutes before delivery in 2024.
    \textbf{(a)} Across the time of delivery, the price level is highest in the evening hours, shows a secondary peak in the morning, and is lowest in the early afternoon.
    \textbf{(b)} Across the day of the week, the price level is high on working days and decreases towards the weekend.
    \textbf{(c)} Across the month of the year, the price level is highest in winter and lower in the other months.}
    \label{fig:seasonability}
\end{figure*}

\begin{figure*}[t]
    \centering
    \hspace*{-1.9mm}
    \includegraphics[width=1.025\textwidth]
    {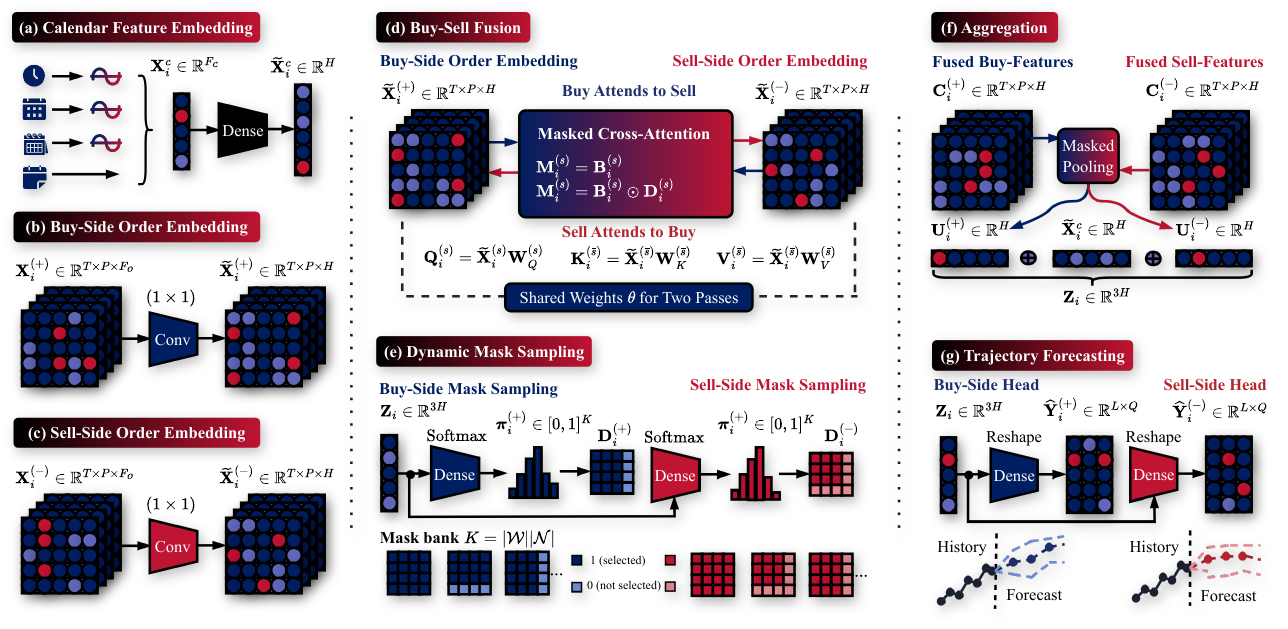}
    \caption{The structure of OrderFusion+. The computation follows the sequence:
\textbf{(a)}  \textbf{(b)}  \textbf{(c)} embed input features
$\rightarrow$ \textbf{(d)} fuse buy and sell features
$\rightarrow$ \textbf{(f)} aggregate  fused features
$\rightarrow$ \textbf{(e)} sample dynamic masks
$\rightarrow$ \textbf{(d)} recompute cross-attention using sampled masks
$\rightarrow$ \textbf{(f)} aggregate updated features
$\rightarrow$ \textbf{(g)} produce probabilistic buy--sell price trajectory forecasts.}
    \label{fig:OrderFusion_Plus}
\end{figure*}

\section{Preliminaries}
\label{sec:preliminary}
\subsection{Continuous Intraday Electricity Market}
\label{sec:continuous-intraday-market}
Intraday markets allow participants to adjust energy positions as delivery
approaches, helping align supply and demand when renewable generation
differs from day-ahead forecasts~\cite{wozabal2020vpp}.
This flexibility becomes increasingly important as the share of intermittent
renewable generation grows.
For 15-min products, continuous intraday
trading \textbf{opens} at 15:00 on the previous day in Germany and many of the other countries covered by EPEX SPOT, and \textbf{closes} at $t_d-\delta_c$ on the current day, where $\delta_c$\footnote{Country-specific
trading cutoffs, in minutes, can be retrieved from the market specifications
available at \url{https://www.epexspot.com/en/downloads}.} is a country-specific
cutoff equal to 30 minutes for Germany, after which trading
continues within the corresponding Transmission System Operator (TSO) control area
until five minutes before delivery.

A defining characteristic of continuous intraday trading is that products with different delivery times can be traded in parallel. For example, as visualized in Fig.~\ref{fig:neighboring_products}, a trader can, at 15:00, simultaneously submit orders for the products delivered at 18:00, 19:00, and 20:00, as well as for any 15-minute product in between or thereafter.
Whereas the non-storability of electricity was generally regarded as a justification for modeling prices at different delivery periods as independent, the increasing influence of battery energy storage system (BESS) arbitrage and demand-side load shifting has caused, and will continue to cause, greater temporal connectedness~\cite{cao2020deep}.
Previous work has shown that modeling dependencies between neighboring
products can improve forecasting performance~\cite{hirsch2024crossproduct}.
Orders submitted at 15:00 for distant delivery (e.g. 20:00) may  provide
additional predictive information by reflecting expectations about market
conditions in the near future.


\subsection{Price Index and Price Trajectory}
\label{sec:price-index-trajectory}

\noindent\textbf{Price Index.}
The Volume-Weighted Average Price (VWAP) is defined as
\begin{equation}
\mathrm{VWAP}=
\frac{\sum_{j\in\mathcal{J}} P_jV_j}
{\sum_{j\in\mathcal{J}} V_j},
\label{eq:price-index}
\end{equation}
where $\mathcal{J}$ denotes the set of aggregated transactions, and
$P_j$ and $V_j$ denote the price and traded volume of transaction $j$.
EPEX SPOT defines $\mathrm{ID}_3$, $\mathrm{ID}_2$, and $\mathrm{ID}_1$
by aggregating transactions from both buy and sell sides,
$\mathcal{S}=\{+,-\}$, over the time window
$\mathcal{T}=[t_d-t_\Delta,t_d-\delta_c]$.
Accordingly,
$\mathcal{J}=\{j:s_j\in\mathcal{S},\,t_j\in\mathcal{T}\}$,
where $t_\Delta=180$, $120$, and $60$ minutes for
$\mathrm{ID}_3$, $\mathrm{ID}_2$, and $\mathrm{ID}_1$, respectively.

\noindent\textbf{Price Trajectory.}
Instead of compressing all transaction prices into a single index, a price
trajectory retains their evolution over time~\cite{pinson2009probabilistic}.
Building on the use of 15-minute VWAP sequences in prior
work~\cite{serafin2022trading}, we compute VWAPs for every 15-minute
interval. As visualized in Fig.~\ref{fig:seasonability}, the aggregated price trajectory exhibits periodic patterns across the time of delivery, the day of the week, and the month of the year. To retain the finer side-specific information, we compute the trajectories of the buy and sell sides separately.
For example, for a forecast produced at $t_d-180$ min covering the trajectory
from $t_d-180$ min to $t_d$, the forecasting targets are
\begin{equation}
  {\mathbf Y}^{(+)} =
  \bigl[
  \mathrm{VWAP}^{(+)}_{t_d-180\,\mathrm{min}:t_d-165\,\mathrm{min}},
  \ldots,
  \mathrm{VWAP}^{(+)}_{t_d-15\,\mathrm{min}:t_d}
  \bigr],
  \label{eq:price-trajectory-buy}
\end{equation}
\begin{equation}
  {\mathbf Y}^{(-)} =
  \bigl[
  \mathrm{VWAP}^{(-)}_{t_d-180\,\mathrm{min}:t_d-165\,\mathrm{min}},
  \ldots,
  \mathrm{VWAP}^{(-)}_{t_d-15\,\mathrm{min}:t_d}
  \bigr].
  \label{eq:price-trajectory-sell}
\end{equation}
As no trades occur during the final 5 minutes before delivery, the final
trajectory interval equivalently contains trades only up to $t_d-5$ min.
The resulting trajectories preserve price trends approaching delivery and
retain differences between the buy and sell sides.

\subsection{OrderFusion}
\label{sec:orderfusion}
OrderFusion is a deep learning model developed to model buy--sell
interactions in the orderbook~\cite{yu2026orderfusion}.
It encodes buy and sell prices and traded volumes
via cross-attention.
Cross-attention produces buy-side representations contextualized by the
sell side and sell-side representations contextualized by the buy side.
Despite its effectiveness in representation learning, OrderFusion ignores neighboring products, and it only forecasts a single price index. This not only overlooks the predictive power contained in neighboring products but also discards side differences and price trajectory dynamics.
Taken together, these limitations motivate us to make an important development to OrderFusion by incorporating neighboring products and producing buy--sell price trajectory forecasts.

\section{OrderFusion+}
\label{sec:model}
We therefore introduce OrderFusion+, an end-to-end model for probabilistic buy--sell
price trajectory forecasting, as illustrated in Fig.~\ref{fig:OrderFusion_Plus}.
For sample $i$, the side-specific
orderbook input is
$\mathbf X_i^{(s)}\in\mathbb R^{T\times P\times F_o}$, where $T$ is the
number of historical timesteps, $P$ contains the target product and its
neighboring products, and $F_o$ is the feature dimension, containing price, traded volume, time position,
and product position. Calendar features are denoted by
$\mathbf X_i^c\in\mathbb R^{F_c}$. The model predicts
$\widehat{\mathbf Y}_i^{(s)}\in\mathbb R^{L\times Q}$, where $L$ is the
forecasting length and $Q$ is the number of quantiles. Throughout this section,
$\mathcal F_{\mathrm{dense}}$ and $\mathcal F_{\mathrm{conv}}$ denote dense and
2D-convolutional layers, respectively. Separate occurrences have independent
parameters unless weight sharing is stated explicitly.

\subsection{Backbone}
\label{sec:backbone}


\noindent \textbf{Calendar Feature Embedding Layer.} As shown in Section~\ref{sec:price-index-trajectory}, intraday prices exhibit periodic patterns. Therefore, the model should  be informed of the position of the delivery in time. We describe time of delivery by its quarter-hour index $q_i\in\{0,\ldots,95\}$,
day of week by $d_i\in\{0,\ldots,6\}$, and month of year by
$m_i\in\{0,\ldots,11\}$. Their periodicity is encoded by
$\psi(x,r)=[\sin(2\pi x/r),\cos(2\pi x/r)]$, while holidays are represented
by the indicator $h_i$:
\begin{equation}
\label{eq:calendar-input}
\mathbf X_i^c=
\left[\psi(q_i,96),\psi(d_i,7),
\psi(m_i,12),h_i\right]^\top\in\mathbb R^{F_c}.
\end{equation}
A dense layer maps the calendar features to the dimension of $H$.
\begin{equation}
\label{eq:calendar-embedding}
\widetilde{\mathbf X}_i^c
=\mathcal F_{\mathrm{dense}}(\mathbf X_i^c)
\in\mathbb R^H,
\end{equation}

\noindent\textbf{Buy--Sell Fusion Layer.}
In continuous trading, buyers often form bids in response to sellers' prices
and volumes, and sellers respond analogously to the buy side. Inspired by this, we use cross-attention
to produce buy representations contextualized by sell information and
sell representations contextualized by buy information. In contrast to
OrderFusion, which uses conventional attention and only considers the
timestep and feature dimension, OrderFusion+ implements customized cross-attention to
model timestep, feature, and product interactions. 

Before feeding into cross-attention, we first map the side-specific input
from $\mathbb R^{T\times P\times F_o}$ to
$\mathbb R^{T\times P\times H}$ 
via a 2D convolutional layer to increase the expressivity of the raw features:
\begin{equation}
\label{eq:side-embedding1}
\widetilde{\mathbf X}_i^{(+)}
=\mathcal F_{\mathrm{conv}}^{(+)}(\mathbf X_i^{(+)})
\in\mathbb R^{T\times P\times H},
\end{equation}
\begin{equation}
\label{eq:side-embedding2}
\widetilde{\mathbf X}_i^{(-)}
=\mathcal F_{\mathrm{conv}}^{(-)}(\mathbf X_i^{(-)})
\in\mathbb R^{T\times P\times H}.
\end{equation}
Its $1\times1$ kernel changes only the feature dimension and leaves the
timestep and product dimensions unchanged. This is intentional, as the
dynamic mask introduced later will remove noisy timesteps and products.

Next, these hidden representations are fed into cross-attention. 
Specifically, 
side $s$ supplies the query, while the opposite side $\bar{s}$ supplies the
key and value:
\begin{equation}
\label{eq:query}
\mathbf Q_i^{(s)}
=\widetilde{\mathbf X}_i^{(s)}\mathbf W_Q^{(s)}.
\end{equation}
\begin{equation}
\label{eq:key}
\mathbf K_i^{(\bar{s})}
=\widetilde{\mathbf X}_i^{(\bar{s})}\mathbf W_K^{(\bar{s})}.
\end{equation}
\begin{equation}
\label{eq:value}
\mathbf V_i^{(\bar{s})}
=\widetilde{\mathbf X}_i^{(\bar{s})}\mathbf W_V^{(\bar{s})}.
\end{equation}
The side-specific contextualized representation is
\begin{equation}
\label{eq:cross-attention}
\mathbf C_i^{(s)}=
\operatorname{Softmax}\!\left(
\frac{\mathbf Q_i^{(s)}(\mathbf K_i^{(\bar{s})})^{\top}}
{\sqrt{d}}
\right)\mathbf V_i^{(\bar{s})}
\in\mathbb R^{T\times P\times H},
\end{equation}
where $d$ is the dimension of one attention head. 

The attention mask $\mathbf M_i^{(s)}$ determines which timestep--product positions are available to the cross-attention layer. Masked key--value positions receive an attention score of $-\infty$ before the softmax operation, while masked query outputs are set to zero. The mask is defined as
\begin{equation}
\mathbf M_i^{(s)}=
\begin{cases}
\mathbf B_i^{(s)}, & \text{initial pass},\\
\mathbf B_i^{(s)}\odot\mathbf D_i^{(s)}, & \text{second pass}.
\end{cases}
\label{eq:effective-mask}
\end{equation}
where $\mathbf B_i^{(s)}\in\{0,1\}^{T\times P}$ denotes the side-specific
missing-value mask, with ``1'' indicating an observed position and ``0''
indicating that no trade occurred in the corresponding trading interval.
In the first pass, only $\mathbf B_i^{(s)}$ is used as the attention mask to
produce the representation used by the \textbf{Dynamic Mask Sampling Layer}.
Once $\mathbf D_i^{(s)}$ has been sampled, the second pass recomputes the
cross-attention representation.
The two passes share all cross-attention weights. Thus, the dynamic mask changes
only the information available to the model and does not introduce additional parameters.

\noindent\textbf{Dynamic Mask Sampling Layer.}


\noindent The optimal historical window and number of neighboring products can change with the market
condition and can differ between the buy and sell sides. Therefore, we  define
the candidate historical windows as
$\mathcal W=\{15,30,60,120,180\}$ minutes and the candidate numbers of
neighboring products as $\mathcal N=\{0,1,2,4,8,12\}$. Their Cartesian product
forms the mask bank
$\mathcal D=\{\mathbf D_1,\ldots,\mathbf D_K\}$, where
$K=|\mathcal W||\mathcal N|$. The mask $\mathbf D_k\in\{0,1\}^{T\times P}$ encodes the prior that older observations and more
distant products should be included only when they provide additional
information. 
Example masks are
\[
\resizebox{\columnwidth}{!}{$\displaystyle
\begin{gathered}
\underbrace{
\begin{bmatrix}
1&0&0&\cdots&0\\
0&0&0&\cdots&0\\
\vdots&\vdots&\vdots&\ddots&\vdots\\
0&0&0&\cdots&0
\end{bmatrix}}_{\text{\textbf{(a)} 15 min, 0 neighbors}}
\qquad
\underbrace{
\begin{bmatrix}
1&1&1&\cdots&1\\
0&0&0&\cdots&0\\
\vdots&\vdots&\vdots&\ddots&\vdots\\
0&0&0&\cdots&0
\end{bmatrix}}_{\text{\textbf{(b)} 15 min, 12 neighbors}}
\qquad
\underbrace{
\begin{bmatrix}
1&1&0&\cdots&0\\
1&1&0&\cdots&0\\
\vdots&\vdots&\vdots&\ddots&\vdots\\
1&1&0&\cdots&0
\end{bmatrix}}_{\text{\textbf{(c)} 180 min, 1 neighbor}}
\\[2ex]
\underbrace{
\begin{bmatrix}
1&1&1&\cdots&1\\
1&1&1&\cdots&1\\
\vdots&\vdots&\vdots&\ddots&\vdots\\
1&1&1&\cdots&1
\end{bmatrix}}_{\text{\textbf{(d)} 180 min, 12 neighbors}}
\qquad
\underbrace{
\begin{bmatrix}
0&0&0&\cdots&0\\
1&1&1&\cdots&1\\
0&0&0&\cdots&0\\
\vdots&\vdots&\vdots&\ddots&\vdots
\end{bmatrix}}_{\text{\textbf{(e)} invalid: timestep}}
\qquad
\underbrace{
\begin{bmatrix}
0&1&0&\cdots&0\\
0&1&0&\cdots&0\\
\vdots&\vdots&\vdots&\ddots&\vdots\\
0&1&0&\cdots&0
\end{bmatrix}}_{\text{\textbf{(f)} invalid: product}}
\end{gathered}
$}\]
Mask~\textbf{(a)} represents a rapidly changing market in which predictive
information is highly concentrated in the latest 15-minute window and neighboring
products provide no information. Masks~\textbf{(b)} and~\textbf{(c)}
may reflect a buyer or seller splitting a large order across trading intervals
or neighboring products to reduce its price impact. Mask~\textbf{(d)} retains
the full historical and product dimensions (target product with 12 neighboring products). Masks~\textbf{(e)} and~\textbf{(f)}
illustrate invalid structures: mask~\textbf{(e)} skips the most recent timestep
while retaining an older one, whereas mask~\textbf{(f)} excludes the target
product while retaining a neighboring product. The masks are broadcast over
the hidden dimension when applied to a $T\times P\times H$ representation.

We denote by $\mathbf Z_i$ the representation obtained from the \textbf{Aggregation
Layer}. We apply side-specific dense layers to assign probabilities to the
mask bank:
\begin{equation}
\label{eq:selector-probability-buy}
\boldsymbol\pi_i^{(+)}
=\operatorname{Softmax}\!\left(
\mathcal F_{\mathrm{dense}}^{(+)}(\mathbf Z_i)\right)
\in[0,1]^K.
\end{equation}
\begin{equation}
\label{eq:selector-probability-sell}
\boldsymbol\pi_i^{(-)}
=\operatorname{Softmax}\!\left(
\mathcal F_{\mathrm{dense}}^{(-)}(\mathbf Z_i)\right)
\in[0,1]^K.
\end{equation}
For each side, the dynamic mask $\mathbf D_i^{(s)}$ is drawn from
$\mathcal D$ according to $\boldsymbol\pi_i^{(s)}$ during training.
Deterministic prediction uses the candidate with the highest probability.

\noindent\textbf{Aggregation Layer.}
The cross-attention outputs are aggregated by a mask-normalized
average:
\begin{equation}
\label{eq:masked-aggregation}
\mathbf U_i^{(s)}=
\frac{
\displaystyle\sum_{t=1}^{T}\sum_{p=1}^{P}
M_{i,t,p}^{(s)}\mathbf C_{i,t,p}^{(s)}
}{
\displaystyle\max\!\left(1,
\sum_{t=1}^{T}\sum_{p=1}^{P}M_{i,t,p}^{(s)}\right)
}
\in\mathbb R^H.
\end{equation}
The maximum prevents division by zero when no valid position is available.
Otherwise, the denominator equals the number of valid positions. Consequently,
discarded positions neither contribute to the representation nor alter its
scale. 

The resulting buy and sell representations are combined with the calendar
embedding:
\begin{equation}
\label{eq:selector-context}
\mathbf Z_i=
\operatorname{concat}\!\left(
\mathbf U_i^{(+)},
\mathbf U_i^{(-)},
\widetilde{\mathbf X}_i^c
\right)
\in\mathbb R^{3H}.
\end{equation}

\subsection{Head}
\label{sec:head}
Two dense layers take the aggregated and concatenated representation $\mathbf Z_i$ as input to generate the side-specific probabilistic
trajectories:
\begin{equation}
\label{eq:forecast-output1}
\widehat{\mathbf Y}_i^{(+)}
=\mathcal F_{\mathrm{dense}}^{(+)}(\mathbf Z_i)
\in\mathbb R^{L\times Q},
\end{equation}
\begin{equation}
\label{eq:forecast-output2}
\widehat{\mathbf Y}_i^{(-)}
=\mathcal F_{\mathrm{dense}}^{(-)}(\mathbf Z_i)
\in\mathbb R^{L\times Q}.
\end{equation}
Here, $L$ is the forecasting length and $Q$ is the number of quantiles. We
define $\mathcal Q=\{0.1,0.5,0.9\}$ to demonstrate functionality.

\subsection{Loss}
\label{sec:loss}
Let $\Omega$ denote the set of observed target positions $(i,s,l)$. Missing
targets are excluded from both the numerator and denominator. Following
\cite{yu2025pricefm}, we jointly estimate all quantiles using the masked
Average Quantile Loss (AQL):
\begin{equation}
\label{eq:trajectory-aql}
\mathcal L_{\mathrm{AQL}}=
\frac{1}{Q|\Omega|}
\sum_{(i,s,l)\in\Omega}\sum_{\tau\in\mathcal Q}
\rho_{\tau}\!\left(
Y_{i,l}^{(s)}-\widehat Y_{i,l,\tau}^{(s)}
\right),
\end{equation}
where the quantile loss is
\begin{equation}
\label{eq:pinball}
\rho_{\tau}(e)=
\begin{cases}
\tau e, & e\geq0,\\
(\tau-1)e, & e<0.
\end{cases}
\end{equation}

\section{Baselines}
\label{sec:baselines}
We compare OrderFusion+ with four persistence baselines, four fully trained models, and four pretrained foundation models. 

\subsection{Persistence}
The persistence baselines
 are strong and
often challenging benchmarks in intraday price forecasting
\cite{janke2019distribution,narajewski2020econometric,lago2021benchmark}.
The first three use the VWAP over the latest 15, 30, and 60 minutes,
respectively, as the point forecast for every step of the future trajectory.
The fourth uses the trajectory of the same delivery product from the
previous day. For each
persistence baseline, empirical residual percentiles from the preceding seven
days provide the probabilistic forecasts.

\subsection{Fully Trained Models}
We additionally compare fully trained models that receive the same inputs as OrderFusion+. Linear Quantile {Regression (LQR) is a common forecasting model in the continuous intraday market~\cite{serafin2022trading,narajewski2020econometric}. It is included to investigate whether a simple linear model is sufficient for price trajectory forecasting.
The Multi-Layer Perceptron (MLP), in contrast, is a non-linear deep learning model. According to the Universal Approximation Theorem (UAT)~\cite{hornik1989multilayer}, a multi-layer neural network can approximate any continuous mapping. However, both baselines simply flatten the inputs and thereby lose the structural prior of the timestep and product dimensions. LSTM and
Transformer are popular temporal models for intraday price forecasting~\cite{kilic2024lstm,yu2026orderfusion}. They retain the timestep dimension, however, neither
represents the product dimension. Therefore, the product features are concatenated
along the feature dimension.}

\subsection{Pretrained Foundation Models}
Recent time-series foundation models are pretrained on large real-world or
synthetic collections spanning domains such as transportation, energy, and
finance, and have shown strong zero-shot forecasting performance in unseen
domains. {We therefore include the latest foundation models, TimesFM~3.0, Chronos~2.0, TabPFN-TS, and Moirai~2.0, as
 baselines
\cite{google2026timesfm3,ansari2025chronos2,hoo2025tabpfnts,liu2025moirai2}. The model information is summarized in Table~\ref{tab:foundation-models}. As these models are pretrained on other large datasets, they are used to examine whether a domain-specific model is still needed.}

\begin{table}[t]
  \caption{Time-series foundation models used for zero-shot forecasting.
  Parameter counts correspond to the evaluated checkpoints.}
  \label{tab:foundation-models}
  \centering
  \renewcommand{\arraystretch}{1.15}

  \resizebox{\columnwidth}{!}{%
  \begin{tabular}{
    *{5}{>{\centering\arraybackslash}p{1.47cm}}
  }
    \toprule
    Model & Parameters & Source & Probabilistic & Date \\
    \midrule
    TimesFM-3
    & 330.7 M
    & \href{https://huggingface.co/google/timesfm-3.0-pytorch}{Google}
    & \checkmark
    & 2026-08-31 \\

    Chronos-2
    & 119.5 M
    & \href{https://huggingface.co/amazon/chronos-2}{Amazon}
    & \checkmark
    & 2025-10-17 \\

    TabPFN-TS
    & 58.0 M
    & \href{https://huggingface.co/Prior-Labs/tabpfn_3}{Prior Labs}
    & \checkmark
    & 2026-05-12 \\

    Moirai 2.0
    & 11.4 M
    & \href{https://huggingface.co/Salesforce/moirai-2.0-R-small}{Salesforce}
    & \checkmark
    & 2025-11-12 \\
    \bottomrule
  \end{tabular}%
  }
\end{table}

\begin{table}[t]
  \caption{Three-fold data splits for training, validation, and testing. The testing collectively covers the full year in 2024.}
  \label{tab:data-splits}
  \centering
  \renewcommand{\arraystretch}{1.15}

  \resizebox{\columnwidth}{!}{%
  \begin{tabular}{
    c
    *{3}{>{\centering\arraybackslash}p{2.1cm}}
  }
    \toprule
    Fold & Training & Validation & Testing \\
    \midrule
    1 & 2022.01--2023.08 & 2023.09--2023.12 & 2024.01--2024.04 \\
    2 & 2022.01--2023.12 & 2024.01--2024.04 & 2024.05--2024.08 \\
    3 & 2022.01--2024.04 & 2024.05--2024.08 & 2024.09--2024.12 \\
    \bottomrule
  \end{tabular}%
  }
\end{table}

\begin{table}[t]
  \centering
  \caption{Testing performance averaged over the three forecasting origins. AQCE and AQCR are reported in percentage (\%). Lower values are better for AQL, AQCE, AQCR, MAE, and RMSE, while higher values are better for $R^2$. The best result is shown in bold, and the second-best is marked with underline.}
  \label{tab:aggregate-comparison}

  \resizebox{\columnwidth}{!}{%
  \begin{tabular}{
    l
    *{6}{>{\centering\arraybackslash}p{.7cm}}
  }
    \toprule
    Model & AQL & AQCE & AQCR & MAE & RMSE & $R^2$ \\
    \midrule
    Persistence-1 & 12.87 & \textbf{0.33} & \textbf{0.00} & 32.23 & 111.25 & 0.23 \\
    \rowcolor{gray!15}
    Persistence-2 & 12.07 & \underline{0.33} & \textbf{0.00} & 30.71 & 105.55 & 0.32 \\
    Persistence-3 & 21.00 & 0.39 & \textbf{0.00} & 52.26 & 159.51 & -1.22 \\
    Persistence-4 & 21.95 & 0.70 & \textbf{0.00} & 56.91 & 165.32 & -0.59 \\
    \midrule
    LQR & 15.85 & 4.19 & 3.09 & 41.24 & 110.80 & 0.26 \\
    MLP & 13.93 & 2.68 & 0.26 & 37.91 & 107.93 & 0.32 \\
    LSTM & 11.79 & 3.58 & 0.06 & 30.81 & 105.21 & 0.35 \\
    \rowcolor{gray!15}
    Transformer & \underline{11.72} & 2.96 & 0.07 & 30.33 & 106.32 & 0.34 \\
    \midrule
    TimesFM 3.0 & 14.85 & 5.74 & \textbf{0.00} & 30.90 & 117.64 & 0.15 \\
    Chronos 2.0 & 19.05 & 4.28 & \textbf{0.00} & 31.17 & 105.25 & 0.33 \\
    \rowcolor{gray!15}
    TabPFN-TS & 12.05 & 8.62 & \textbf{0.00} & \underline{28.63} & \underline{102.86} & \underline{0.36} \\
    Moirai 2.0 & 17.66 & 3.51 & \underline{0.01} & 36.88 & 112.62 & 0.22 \\
    \midrule

    \rowcolor{gray!15}
    OrderFusion+ & \textbf{10.62} & 2.29 & \textbf{0.00} & \textbf{28.04} & \textbf{101.76} & \textbf{0.39} \\
    \bottomrule
  \end{tabular}%
  }
\end{table}

\begin{figure}[t]
    \centering
    \hspace*{-8mm}
\includegraphics[width=0.42\textwidth]
    {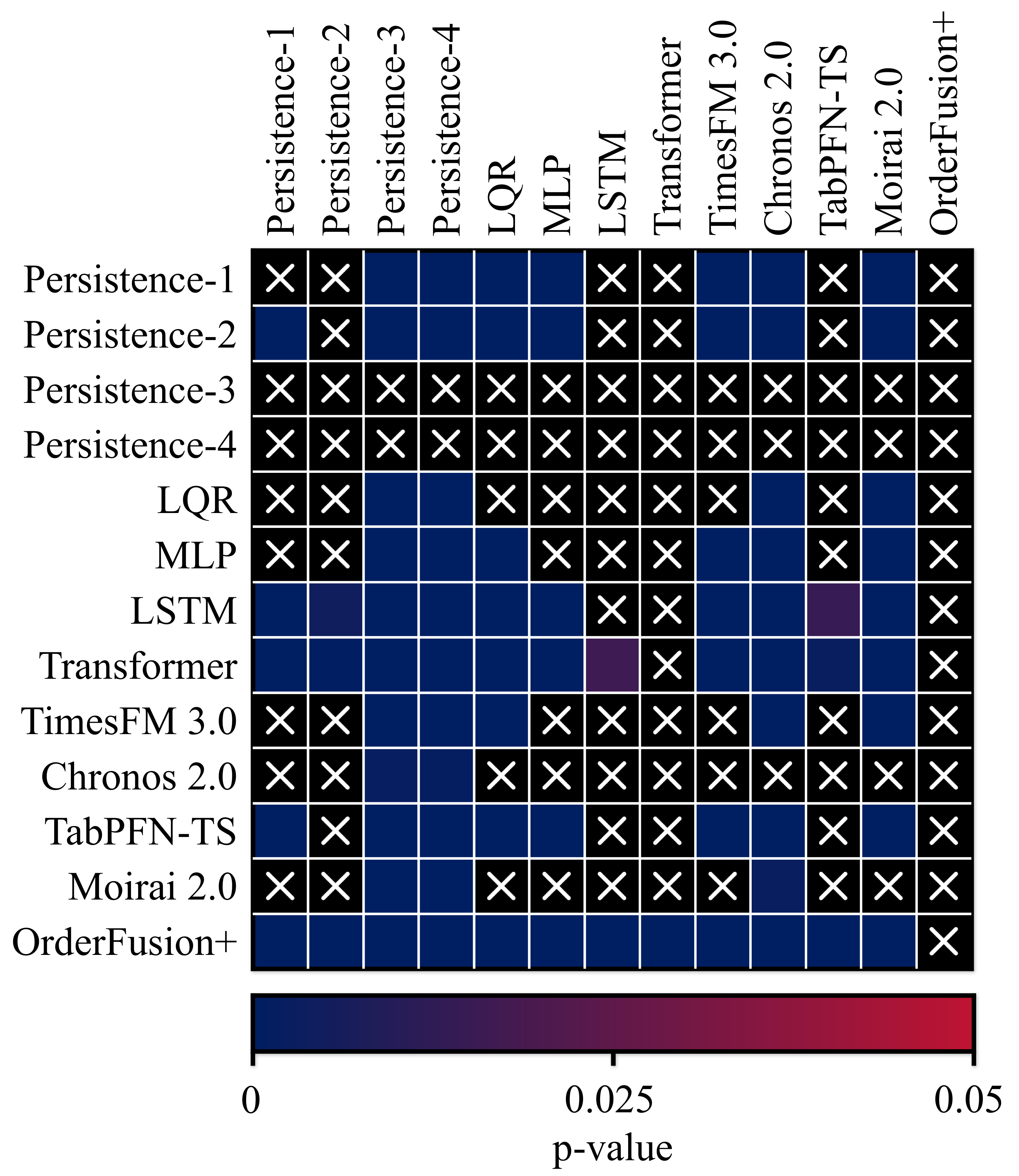}
    \caption{One-sided DM $p$-values of the AQL loss differential. The cell in row $A$ and column $B$ gives the $p$-value of the hypothesis that model $A$ is more accurate than model $B$; a colored cell indicates $p<0.05$ (model $A$ significantly beats model $B$), and a cross indicates $p\geq0.05$.}
    \label{fig:DM_test}
\end{figure}

\section{Evaluation Metrics}
\label{sec:metrics}

For probabilistic forecasting, we use Average Quantile Loss (AQL),
Average Quantile Coverage Error (AQCE), and Average Quantile Crossing Rate
(AQCR), following prior work~\cite{yu2025pricefm, yu2026orderfusion}.
For point forecasting, we use Mean Absolute Error (MAE), Root Mean Squared
Error (RMSE), and the coefficient of determination ($R^2$).
To test whether one forecast is significantly more accurate than another,
we apply a one-sided Diebold--Mariano (DM) test~\cite{diebold1995comparing}.

{AQCE} is the mean absolute deviation of the empirical coverage
from the nominal quantile level:
\begin{equation}
  \mathrm{AQCE}
  =\frac{1}{Q}\sum_{\tau\in\mathcal{Q}}
  \left|\widehat{c}_\tau-\tau\right|,
\end{equation}
where
\begin{equation}
  \widehat{c}_\tau
  =\frac{1}{|\Omega|}
  \sum_{(i,s,l)\in\Omega}
  \mathbb{I}\!\left\{
  Y_{i,l}^{(s)}
  \leq
  \widehat{Y}_{i,l,\tau}^{(s)}
  \right\},
\end{equation}
and $\mathbb{I}\{\cdot\}$ denotes the indicator function.

{AQCR} is the fraction of adjacent predicted quantiles that violate
their ordering:
\begin{equation}
  \mathrm{AQCR}
  =\frac{1}{|\Omega|(Q-1)}
  \sum_{(i,s,l)\in\Omega}
  \sum_{q=1}^{Q-1}
  \mathbb{I}\!\left\{
  \widehat{Y}_{i,l,\tau_q}^{(s)}
  >
  \widehat{Y}_{i,l,\tau_{q+1}}^{(s)}
  \right\},
\end{equation}
where $\tau_1<\cdots<\tau_Q$ are the ordered quantile levels in
$\mathcal{Q}$.




{The DM Test} examines whether one model is statistically more accurate
than another. For two models $A$ and $B$, it tests the null hypothesis of
equal predictive accuracy against the one-sided alternative that model $A$
is more accurate. The test is based on the chronologically ordered
per-delivery AQL differential, using a heteroskedasticity- and
autocorrelation-consistent variance estimate and the small-sample correction
of \citet{harvey1997testing}. A $p$-value below 0.05 indicates that model
$A$ is significantly more accurate than model $B$.

\section{Experiments}
\label{sec:experiment}
\subsection{Experimental Settings}
\label{sec:experimental-settings}
{We use commercial German orderbook data, as the German market is the largest in Europe~\cite{semmelmann2026quantile}. Note that, owing to the market coupling mechanism, the German orderbook also contains orders from traders in other regions, such as France, Austria, the Netherlands, and Norway~\cite{cary2026efficiency}.
The 15-minute products are the forecasting targets, as they are the most volatile and thus the most interesting to traders~\cite{yu2026orderbookfeatures}.
A rolling horizon with a 15-minute step and three
train--validation--test folds are applied, as summarized in Table~\ref{tab:data-splits}.}
Both the historical window and neighboring-product coverage span 180 minutes.
We evaluate three prediction origins: $-180$, $-120$, and
$-60$ minutes relative to delivery.
The three origins correspond to forecasting lengths of 180, 120,
and 60 minutes, respectively.

\begin{figure*}[t]
    \centering
\includegraphics[width=0.96\textwidth]
    {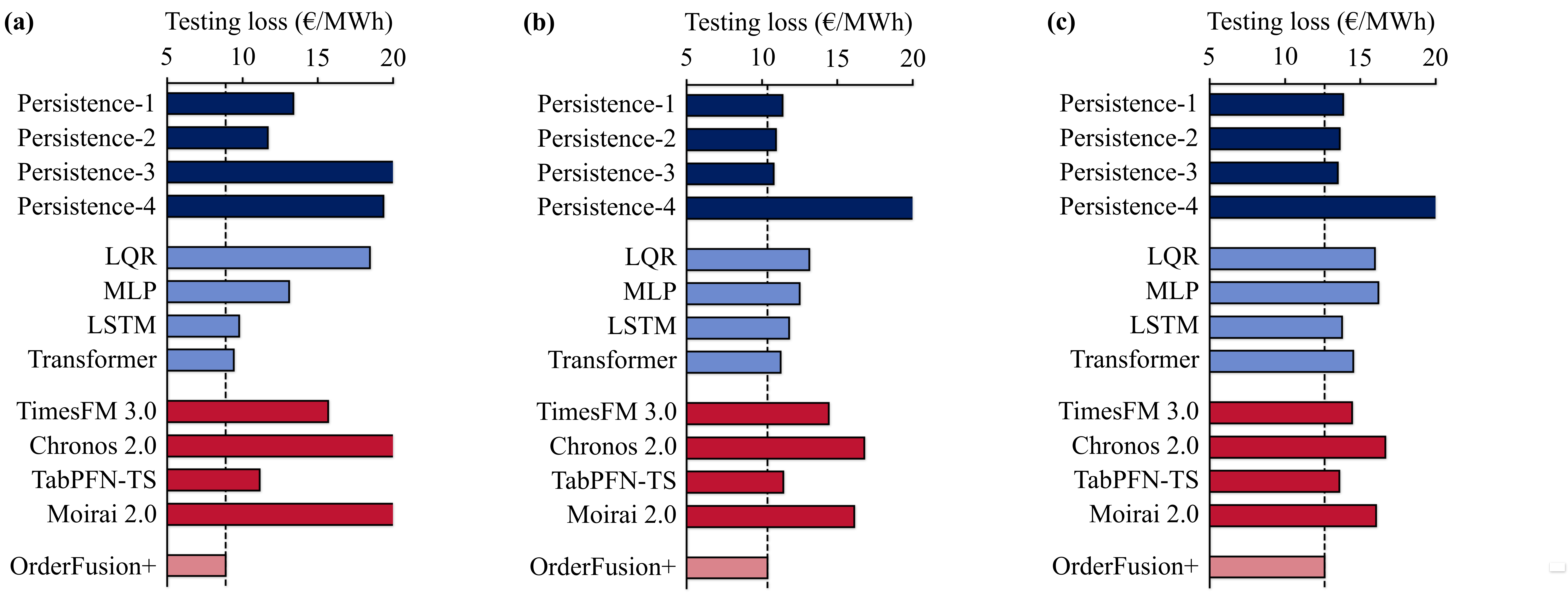}
    \caption{Testing AQL per forecasting origin.
    \textbf{(a)} Origin $=-180$ min. \textbf{(b)} Origin $=-120$ min. \textbf{(c)} Origin $=-60$ min. The testing loss range is constrained within 20 for better visualization. OrderFusion+ outperforms all other models at every origin.}
    \label{fig:Per_Origin_AQL}
\end{figure*}

\begin{figure*}[t]
    \centering
\includegraphics[width=0.96\textwidth]
    {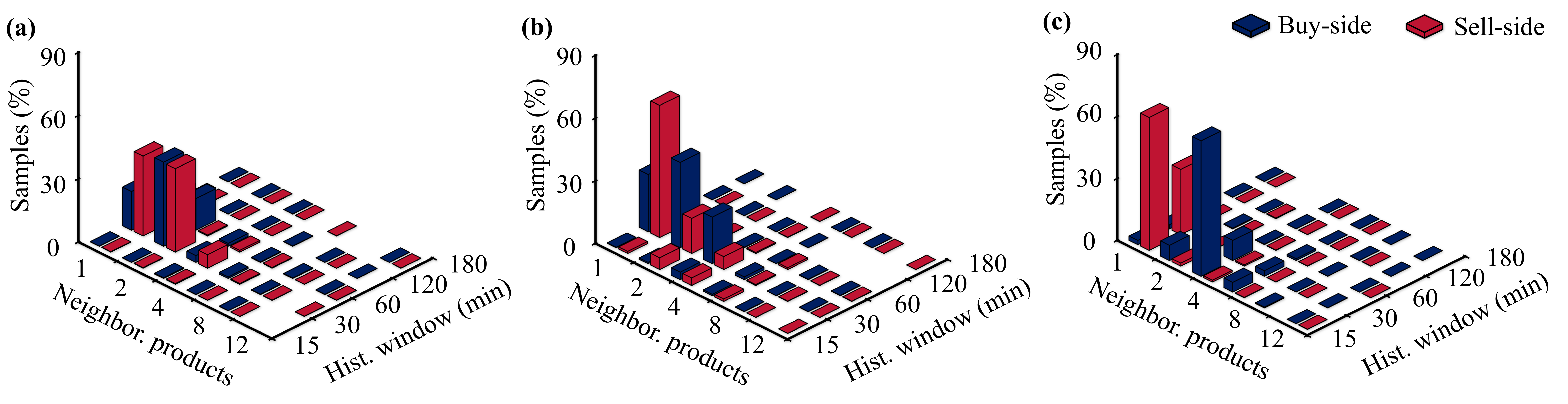}
    \caption{Distribution of the sampled dynamic masks over all testing samples.
    \textbf{(a)} Origin $=-180$ min. \textbf{(b)} Origin $=-120$ min. \textbf{(c)} Origin $=-60$ min. The distribution varies across origins and shifts towards shorter historical windows as delivery approaches.}
    \label{fig:mask_distribution}
\end{figure*}

\subsection{Experimental Results}
\label{sec:experimental-results}
\textbf{Model Comparison.}
Table~\ref{tab:aggregate-comparison} reports the testing performance averaged over the three forecasting origins. Overall, OrderFusion+ outperforms all baselines in AQL, MAE, RMSE, and $R^2$, and produces no quantile crossing, as indicated by an AQCR of 0\%. The DM $p$-values reported in Fig.~\ref{fig:DM_test} confirm that this improvement is statistically significant with respect to every baseline.
In AQCE, OrderFusion+ is slightly worse than the persistence baselines but achieves the lowest value among the non-persistence baselines. The per-origin AQL comparison in Fig.~\ref{fig:Per_Origin_AQL} confirms that OrderFusion+ achieves the best performance at every origin, demonstrating robust performance.

{Among the persistence baselines, Persistence-2 achieves the lowest AQL, MAE, and RMSE and the highest $R^2$, showing that the latest 30-minute VWAP is a strong market indicator. A shorter window (15 min) or a longer window (60 min) degrades the performance. In particular, Persistence-4, which uses the data from the previous day, performs worst among the four persistence baselines.}

{Among the fully trained baselines, LQR performs worst, indicating that the mapping from inputs to outputs is strongly non-linear. MLP is the second worst, which is expected, as it loses the timestep and product priors. LSTM and Transformer perform similarly, as both retain the timestep dimension. However, their AQL is approximately 9.4\% higher than that of OrderFusion+, indicating that preserving the structural prior is critical for intraday price forecasting.}

Among the pretrained foundation models, TabPFN-TS consistently outperforms the other foundation-model baselines. 
{However, TabPFN-TS does not beat Persistence-2 significantly in AQL, as shown in Fig.~\ref{fig:DM_test}, and produces large coverage error of 8.62\%, as shown in Table~\ref{tab:aggregate-comparison}. This implies that generic pretrained foundation models are insufficient for probabilistic intraday price trajectory forecasting. Nevertheless, its point metrics, including MAE, RMSE, and $R^2$, are close to those of OrderFusion+, although still worse. Thus, when the orderbook training data are too limited to train a domain-specific model, TabPFN-TS is an appropriate zero-shot alternative for intraday price trajectory forecasting.}

\textbf{Dynamic Market Condition.}
We investigate the distribution of the sampled dynamic masks across all testing samples. As illustrated in Fig.~\ref{fig:mask_distribution}, the model samples masks dynamically across forecasting origins and market sides, indicating that the relevant market condition is not static.

{In particular, as the forecasting origin moves from $-180$ to $-60$ minutes, the most frequently chosen historical window of both sides shortens from 30 to 15 minutes. This indicates that, as delivery approaches, the predictive information is increasingly concentrated in the most recent transactions, and the model therefore prioritizes more recent information. Aggregated over the three forecasting origins, the most frequently chosen historical window is 30 minutes (63\% of all samples), and the most frequently chosen number of neighboring products is zero (47\%), which is consistent with the competitive performance of Persistence-2.}

Moreover, the buy and sell sides sample masks asymmetrically. A possible explanation is that the two sides exhibit different conditional persistence, liquidity, or recovery dynamics~\cite{hall2007buy,farmer2006efficiency,toth2015persistent}. 
For example, a buyer or seller might distribute a large order along the time or product axis to reduce its price impact.
Therefore, a static model with a fixed historical window and a fixed number of neighboring products, which implicitly assumes an unchanged market condition, is insufficient to capture such dynamics.

Furthermore, cross-attention allows information selected on either side to pass to the other side. Consequently, different side-specific mask allocations may be functionally equivalent. Therefore, the side-specific distributions should not be interpreted as a fixed market conclusion. Instead, one fitted model provides one perspective on the market, consistent with the Rashomon effect\footnote{The Rashomon effect describes the situation in which different model behaviors achieve comparable predictive performance.}~\cite{breiman2001twocultures}. Appendix~\ref{app:rashomon-effect} further discusses the Rashomon effect.

\section{Conclusion}
\label{sec:conclusion}

In this paper, we develop OrderFusion+, a novel deep learning model for probabilistic buy--sell price trajectory forecasting. From the model comparison, we show that OrderFusion+ achieves competitive forecasting performance. We also find that the pretrained foundation models do not outperform a simple persistence baseline significantly, suggesting that generic foundation models remain insufficient for probabilistic intraday price trajectory forecasting. Through the dynamic mask design, we show that the historical window and the number of neighboring products should not be fixed. The resulting mask distributions indicate that the continuous intraday market is dynamic and that the relevance of historical and neighboring-product information changes as delivery approaches. By open-sourcing OrderFusion+, we contribute to the energy forecasting community and provide a basis for developing more advanced trading strategies that incorporate probabilistic  price trajectories.

Several limitations remain. First, the design of the mask bank is empirical, and a finer candidate grid could potentially improve forecasting performance. Second, although prior work has shown that intraday prices reflect macro-level features such as renewable generation and load, it remains unclear whether they also fully reflect weather forecast information. Incorporating additional weather-related features could further improve forecasting performance~\cite{semmelmann2026quantile}. Finally, we evaluate OrderFusion+ only on the German market. Future work should examine its performance in other markets.

\ifusebibliography
  \bibliographystyle{ACM-Reference-Format-citation-order}
  \bibliography{references}
\fi

\appendix
\section{Data Source}
\label{app:data-source}
Orderbook data can be purchased from EPEX SPOT through the
\href{https://webshop.eex-group.com/epex-spot-public-market-data}{EEX Group webshop},
which offers several data types.
For example, the German ``Continuous Anonymous Orders History'' dataset is
listed at \texteuro{}3,900 per calendar year for internal use, equivalent to
\texteuro{}325 per month.


\begin{figure*}[t]
    \centering
\includegraphics[width=0.96\textwidth]
    {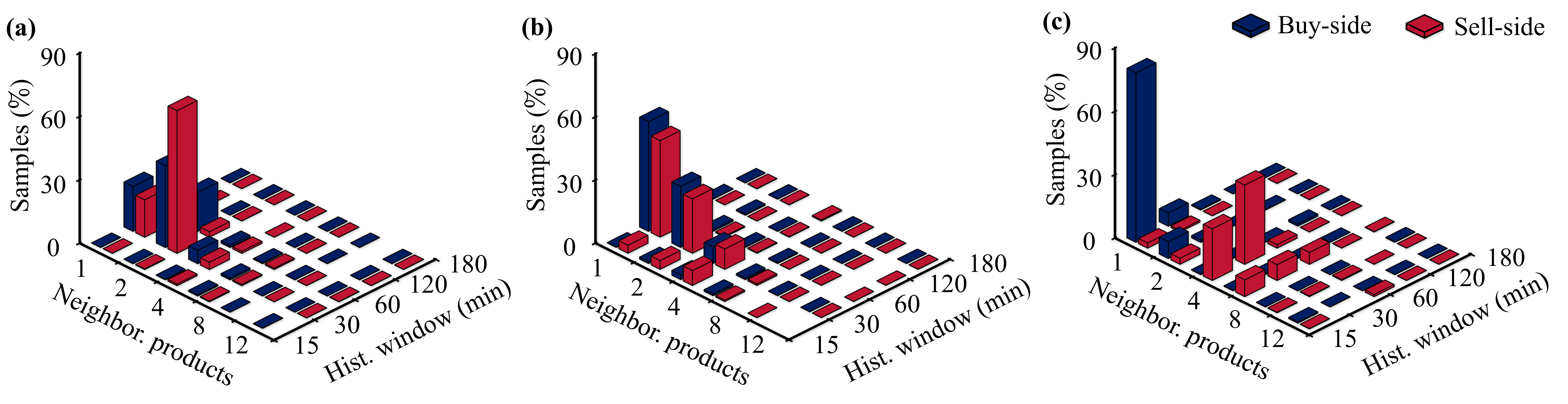}
    \caption{Distribution of the sampled dynamic masks over all testing samples for OrderFusion+$'$ (another random seed).
    \textbf{(a)} Origin $=-180$ min. \textbf{(b)} Origin $=-120$ min. \textbf{(c)} Origin $=-60$ min.}
    \label{fig:mask_distribution_seed42}
\end{figure*}

\section{Rashomon Effect}
\label{app:rashomon-effect}

{We train OrderFusion+ with the same protocol using another random seed, denoted OrderFusion+$'$. As seen from Fig.~\ref{fig:mask_distribution_seed42}, its mask distribution differs from that of OrderFusion+ in Fig.~\ref{fig:mask_distribution}. However, the testing AQL of the two models does not differ significantly (10.62 versus 10.70), as indicated by $p$-value $>0.05$. This corresponds to the Rashomon effect in the machine learning context~\cite{breiman2001twocultures}.

In detail, when both market sides are considered jointly, moving the forecasting origin from $-180$ to $-60$ minutes shortens the most frequently selected historical window from 30 to 15 minutes. This additional seed confirms the previous pattern. Aggregated across the three forecasting origins, the most frequently selected historical window is 30 minutes, accounting for approximately 65\% of all samples, which is close to the 63\% observed for the previous seed. This consistency provides further evidence for why Persistence-2, which uses information from the latest 30 minutes, constitutes a strong baseline.
}


\section{Implementation and Engineering Details}
\label{app:implementation-details}


{For reproducibility, the detailed settings of OrderFusion+ are listed in Table~\ref{tab:orderfusion-implementation}, and the hyperparameter search spaces of the fully trained neural models are listed in Table~\ref{tab:baseline-hyperparameters}. For all fully trained models, we use a sufficiently long training budget of 350 epochs to ensure that the lowest validation loss is always reached before the last epoch, and the model weights with the lowest validation loss are used for the testing evaluation. We use a batch size of 4,096 to utilize the GPU efficiently, the Adam optimizer, and a learning rate of $10^{-3}$. The standard scaler is used for all models.} 

 



\begin{table}[t]
  \caption{Full model configurations of OrderFusion+. The weights of the cross-attention layer are shared for two passes. }
  \label{tab:orderfusion-implementation}
  \centering
  \renewcommand{\arraystretch}{1.15}

  \resizebox{\columnwidth}{!}{%
  \begin{tabular}{
    >{\raggedright\arraybackslash}p{4.5cm}
    >{\raggedright\arraybackslash}p{3.5cm}
  }
    \toprule
    Setting & Value \\
    \midrule
    Aggregation interval & 15 minutes \\
    Historical window & 180 minutes \\
    Number of neighbors & 12 \\
    Hidden dimension & 36 \\
    Cross-attention heads & 2 \\
    Dynamic masks & 30 \\
    Activation & Swish \\
    Quantiles & $\{0.1,0.5,0.9\}$ \\
    \bottomrule
  \end{tabular}%
  }
\end{table}

\begin{table}[t]
  \caption{Hyperparameter search spaces of fully trained neural networks.
  The bold values indicate the selected settings.}
  \label{tab:baseline-hyperparameters}
  \centering
  \renewcommand{\arraystretch}{1.12}

  \resizebox{\columnwidth}{!}{%
  \begin{tabular}{
    >{\raggedright\arraybackslash}p{3.5cm}
    >{\raggedright\arraybackslash}p{4.5cm}
  }
    \toprule
    Model & Hyperparameter search space \\
    \midrule

    MLP
    & Hidden dimension $=\{32,\mathbf{64},128\}$ \\
    & Layers $=\{2,\mathbf{3},4\}$ \\
    & Activation $=\{\textbf{ReLU},\mathrm{Swish}\}$ \\
    & Dropout $=\{\mathbf{0},0.1,0.2\}$ \\
    \addlinespace

    LSTM
    & Hidden dimension $=\{32,\mathbf{64},128\}$ \\
    & Layers $=\{\mathbf{2}, 3, 4\}$ \\
    & Activation $=\{\textbf{ReLU},\mathrm{Swish}\}$\\
    & Dropout $=\{0,\mathbf{0.1},0.2\}$ \\
    \addlinespace

    Transformer
    & Hidden dimension $=\{32,\mathbf{64},128\}$ \\
    & Layers $=\{2,\mathbf{3},4\}$ \\
    & Attention heads $=\{1,\mathbf{2},4\}$ \\
    & Activation $=\{\mathrm{ReLU},\mathrm{\textbf{Swish}}\}$\\
    \addlinespace

    OrderFusion+
    & Hidden dimension $=\{18,\mathbf{36}, 72\}$ \\
    & Attention heads $=\{1,\mathbf{2},4\}$ \\
    & Activation $=\{\mathrm{ReLU},\mathrm{\textbf{Swish}}\}$\\

    \bottomrule
  \end{tabular}%
  }
\end{table}

\section{Hardware and Computation}
\label{app:hardware-computation}

All experiments were conducted in Google Colab using one NVIDIA A100 GPU. The implementation uses CUDA-accelerated PyTorch for the
full-grid cross-attention and model optimization. Training one model required
approximately 25--35 minutes. OrderFusion+ is a small model with only 41,124, 38,508, and 35,892 trainable parameters for the $-180$, $-120$, and $-60$ min origins, respectively. The forecasting length changes the parameter count of the forecasting head.
Inference requires
less than one second, which makes the model suitable for real-time applications.

\end{document}